\documentclass[12pt,preprint]{aastex}

\usepackage{amssymb}
\usepackage{lscape}
\usepackage{graphics}
\usepackage{comment}
\usepackage{subfigure}

\newcommand{\eq}{\begin{equation}}
\newcommand{\ee}{\end{equation}}

\newcommand{\msun}{$M_\odot$}

\begin{document}
\title{On the Coagulation and Size Distribution of Pressure Confined
Cores}
\author{ Xu Huang\altaffilmark{1,2}\email{xuhuang@princeton.edu}, 
Tingtao Zhou\altaffilmark{1},
\& D.N.C. Lin\altaffilmark{1,3}}
\affil{$^1$Kavli Institute for Astronomy \& Astrophysics and
School of Physics, Peking University, Beijing China}
\affil{$^2$Dept of Astrophysical Sciences, Peyton Hall, 4 Ivy Lane,
Princeton University, Princeton, NJ 08540, USA}
\affil{$^3$UCO/Lick Observatory, University of California, Santa
Cruz, California, 95064}

\begin{abstract}
Observations of the Pipe Nebula have led to the discovery of dense
starless cores. The mass of most cores is too small for their self 
gravity to hold them together. Instead, they are thought to be pressure 
confined. The observed dense cores' mass function (CMF) matches well 
with the initial mass function (IMF) of stars in young clusters. 
Similar CMF's are observed in other star forming regions such as
the Aquila Nebula, albeit with some dispersion. The shape of 
these CMF provides important clues to the competing physical 
processes which lead to star formation and its feedback on the 
interstellar media. In this paper, we investigate the 
dynamical origin of the the mass function of starless cores which are 
confined by a warm, less dense medium. In order to follow the evolution 
of the CMF, we construct a numerical method to consider the coagulation 
between the cold cores and their ablation due to Kelvin-Helmholtz 
instability induced by their relative motion through the warm medium. 
We are able to reproduce the observed CMF among the starless cores in 
the Pipe nebula. Our results indicate that in environment similar to 
the Pipe nebula: 1) before the onset of their gravitational collapse, 
the mass distribution of the progenitor cores is similar to that of 
the young stars, 2) the observed CMF is a robust consequence of dynamical 
equilibrium between the coagulation and ablation of cores, and 3) a break 
in the slope of the CMF is due to the enhancement of collisional cross section 
and suppression of ablation for cores with masses larger than the 
cores' Bonnor-Ebert mass.
\end{abstract}


\section{Introduction}
Dense molecular cores are long known to be embedded in molecular clouds. 
These cores are particularly interesting in the context of star 
formation because they are thought to be the progenitors of individual 
stars or small stellar systems. Recent infrared measurements of dust 
extinction, CO and NH$_3$ maps provide means for observers to 
systematically and quantitatively analyze the physical properties of 
these cores. Among the first studies of this kind is that carried out by 
\citet{Alves2007} on the Pipe nebula. These observations revealed a 
population of dense, starless cloud cores. The similarity between the 
density, temperature, and thermal pressure of cores at different 
locations in the nebula suggests that they are confined by a global 
external pressure, although the most massive cores have sufficient self 
gravity to hold themselves together. Prestellar cores are also found in 
the Aquila Nebula, Polaris Nebula and Ophiuchus main cloud 
\citep{Konyves2010,Andre2010,Andre2007}, though there exists 
some diversities in the environment and the observed mass distribution. 

One scenario for the confinement of these cores in Pipe is the possible 
existence of a three-phase medium which includes 1) a hot component 
ionized by the nearby B2 IV $\beta$ Cephei star, $\theta$ Oph, 2) a 
cold atomic gas ($\sim$100K, we refer to it as warm medium to distinguish 
with the molecular phase) , and 3) the cold molecular cores ($\sim$10K) 
\citep{GritschnederLin2012}.
We associate this multi-phase medium in the Pipe nebula with the 
byproduct of a thermal instability which is excited when thermal energy 
in a slightly cooler region is radiated away faster than its 
surrounding \citep{Field1965}. A quasi hydrostatic pressure balance is 
maintained over regions smaller than the Field length scale for sound 
waves to travel within the instantaneous cooling time. These regions 
undergo isobaric cooling such that their density contrast to the 
background grows in time. Meanwhile the Field length declines with 
decreases in both the sound speed and cooling timescale. Eventually, 
density difference between the cold regions and the medium which 
engulf them either become nonlinear or reach an asymptotic limit over 
regions larger than the shrinking Field length \citep{Burkert2000}.

Thermal instability is determined by local conditions and its growth 
rate does not depend on the wavelength. Nevertheless, the initial 
perturbation spectrum does determine the range of length scale over 
which the density contrast may become nonlinear during the onset of 
thermal instability. The interstellar medium is turbulent and attains 
a power-law perturbation spectrum, similar to the Kolmogorov law, over 
an extended range of lengthscales \citep{Lazarian2009}. However, the 
cores' mass function (CMF) in the Pipe nebula appears to resemble more 
closely with the stellar Initial Mass Function (IMF) 
\citep{Alves2007,Salpeter1955,Kroupa2002} than with the single power-law 
spectrum of the interstellar medium. The similarity between the 
CMF and IMF has also been observed in the Aquila Nebula and Ophiuchus main 
cloud \citep{Konyves2010, Andre2007}. Palaris is an exception of this 
resemblance \citep{Andre2010}. 
The relationship between the CMF and IMF has been investigated by many 
authors using different approaches 
\citep{Smith2009,DibShadmehri2010,Klessen2010,Anathpindika2011}. 
While there remains diverse opinions on which physical processes affect 
the evolution of the CMF and the transition to the IMF, it is generally 
recognized that such a process critically determines star formation rate 
and efficiency.

Many past discussions have been focused on the origin of the shape 
of CMF (see \citet{Padoan2002,Hennebelle2008,Smith2009,
Offner2008,DibBrandenburg2008} 
etc), which may be related to the turbulent and self-gravitating flows 
\citep{Mckee2007}. Such flows are complex because nonlinear structure of 
multi phase medium extends over a large dynamical range, from the inertial 
to dissipative scales, throughout the star forming region. Several 
numerical methods, including different versions of the adaptive mesh 
refinement (AMR) scheme, have been constructed to simulate the emergence 
of bound and unbound clusters of cores in a turbulent environment. However, 
none of these attempts can fully resolve cloud complexes in which many core 
coexist with their confining medium. In addition, these simulations 
are time consuming and have been carried out for limited sets of model 
parameters. It is not clear whether they depend on the assumed 
history of the system or are applicable in general.

In this paper, we aim to explain the present-day CMF observed in the Pipe 
nebula, which is just prior to the onset of star formation. 
We suggest that the characteristic CMF spectrum is established through 
the coalescence and ablation of cores rather than precipitated 
directly from a turbulent gas cloud through thermal instability. In \S2, 
we construct a coagulation equation which takes into account various 
relevant physical processes. We design a numerical scheme (presented in 
the appendix) to efficiently solve this integral differential equation. 
This method allows us to follow the cores and the surrounding medium as 
two separate, but coexisting, fluids. We describe our detailed modeling 
based on the properties relevant to the Pipe nebula in \S3. In \S4, we 
show that we can reproduce the observed CMF, including the observed 
peak of mass distribution which occurs in a mass range 3 times heavier 
than that for IMF. Finally, we summarize our findings and discuss their 
implication in \S5.

\section{Important Processes in the Two Phase Media}

We assume all but the largest cloud cores are confined by the pressure of 
an external hot and warm medium. For simplicity, we do not distinguish 
between contributions from the hot and warm medium. In this paper, we 
are mainly concerned with the interaction between the cloud cores and 
the inter-core gas. In this context, we take into account two main 
physics processes: a) close-encounters and collisions between the cold 
cores, due to the velocity dispersion between the cores and b) the dynamic 
interaction between the cold cores and the background gas due to the 
drag on the cores by the inter-core gas. Process a) leads to the 
coagulation between the cores, which induces a shift in the mass 
distribution towards higher masses. Process b) leads to 
ablation of the cores, due to the Kelvin-Holmholz 
instability which causes the mass distribution to evolve towards the 
lower masses. In this section we will provide the basic equations for 
the two processes and order of magnitude estimation of the timescale. In 
addition, we briefly discuss the contribution of conduction at the end 
of this section.

\subsection{Coagulation equation}
\label{coag}

Since we are primarily interested in the bulk properties rather than 
the detailed phase space distribution of the cloud population, we adopt 
a conventional approach to examine the evolution of the distribution 
function
\eq
f(m, t) = \frac{dN (m, t)}{dm}
\ee
where $N (m, t)$ is the number of cores in the mass range $m$ and $m+dm$ 
at a time $t$.  We define that the minimum and maximum masses to be 
$m_{\mu}$ and $m_{\rm max}$ respectively. The rate of change $d (dN/dm)/dt$ 
can be expressed in term of a general coagulation equation 
(see \citep{Murray1996,Shadmehri2004,Dib2007}) such that
\eq
\left(\frac{df(m)}{dt}\right)_{\rm coag} =
\frac{1}{2}\int_{m_{min}}^{m-m_{min}}f(m')f(m-m')G(m,m-m')\,dm'
\ee
\begin{displaymath}
-\int_{m_{\rm min}}^{m_{\rm max}}f(m)f(m')G(m,m')\,dm'.
\label{eq:coag}
\end{displaymath}

For computational simplicity, we assume there is no mass loss during 
each collision. We also assume the physical condition of the inter-core 
gas, cores' mass and kinematic distribution are homogenous and isotropic. 
These assumptions can be relaxed in a series global simulations to be 
presented in the future. We adopt an average velocity dispersion 
$\sigma$ for the cores, which is observed to be 
$\sigma \sim 1$ km\,s$^{-1}$ in Pipe nebula \citep{Onishi1999}. 
Hydrodynamic simulations study the kinematics of molecular cloud cores in 
the presence of turbulence also suggest similar values for the velocity 
dispersion between cores \citep{Offner2008}.
Considering gravitational focusing for massive cores, we adopt a 
collisional kernel
\begin{displaymath}
G(m,m')=(1+\Theta) \pi (R_c(m)+R_c(m'))^2\sigma
\end{displaymath}
where
$\Theta=\frac{c_s^2}{2\sigma^2}\left(\frac{m^{2/3}+m'^{2/3}}{m_{\rm BE}^{2/3}}\right)$,
following equation 7.195 in \citet{Binney&Tremaine2008}.
$M_{\rm BE}$ is the Bonnor-Ebert mass for the cores:
\eq
M_{\rm BE}=1.15 M_{\odot}(\frac{c_s}{0.2{\rm km s^{-1}}})
(\frac{P_{\rm ext}/\kappa}{10^5 {\rm K cm^{-3}}})^{-0.5}
\label{eq: mbe}
\ee
where $P_{\rm ext}$ is the external pressure of the inter-core gas and 
$\kappa$ is the Boltzman constant \citep{Bonnor1956, Ebert1955}. 

Cores gain a significant fraction of their own mass through 
cohesive collisions mostly with cores of comparable sizes with 
a top heavy distribution of CMF. The characteristic growth time scale 
for cores to double their masses is $\tau_{\rm coag} =\lambda / \sigma$ 
where the cores' mean free path $\lambda=1/(\pi R_c(m)^2 n_c)$. 
$n_c$ is number density of cores with sizes $R_c(m)$ and mass $m$. 
The number of cores decreases with merger events. 

Under these conditions, for a typical 1 $m_\odot$ core, we estimate
\eq
\tau_{\rm coag}\sim\frac{1}{{\pi}R_c^2n_{c}\sigma}
= 1.3\ {\rm Myr}\left(\frac{R_c}{0.1\ {\rm pc}}\right)
\left(\frac{f_f}{0.1}\right)^{-1}
\left(\frac{\sigma}{\ {\rm km\,s^{-1}}}\right)^{-1}.
\label{eq:taucoag}
\ee
In the above estimate, we replace $n_c$ with $3 f_f / 4 \pi R_c^3$, 
where the magnitude of the volume filling factor $f_f$ can be 
obtained directly from the observed area filling factor.

\subsection{Core ablation as a result of the 
Kelvin-Helmholtz instability}

The intrinsic ablation of the cores also contributes 
to the coagulation equation. When cores attain a relative velocity with 
respect to the inter-core gas, they encounter a drag force. The shear at 
the gas-core interface leads to the excitation of a Kelvin-Helmholtz 
instability \citep{MurrayLin2004}. Within the growth time scale 
$\tau_{\rm KH}$ for the longest unstable wavelength (a significant 
fraction of $R_c(m)$), about half of the core mass break up into smaller 
cores \citep{Murray1993}. Cores' mass loss rate is 
$\partial m / \partial t \simeq -m/\tau_{\rm KH}$, and the mass at time 
$t$ due to ablation alone can be approximated by 
$m{\sim}m(t_0)\exp(t_0-t)/\tau_{\rm KH}$. For a density contrast between 
the cores and the diffuse background, $D_\rho$,
\eq
\tau_{\rm KH}=\left(\frac{\sigma}{R_c(m)D_{\rho}^\frac{1}{2}}\right)^{-1}
=1\ {\rm Myr}\left(\frac{\sigma}{\rm km s^{-1}}\right)^{-1}\left(\frac{R_c(m)}{0.1\ {\rm pc}}\right)\left(\frac{D_{\rho}}{100}\right)^{1/2}
\label{eq:taukh1}
\ee

In the above expression, we assume the relative velocity between the 
cores and the gas $v_r$ to be comparable to the cores' velocity dispersion 
$\sigma$. It is possible that the turbulence speed of the gas $v_g$ is 
directly determined by some other stirring mechanism such as external 
magnetic fields \citep{LazarianVishniac99}. On the turn-over time scale 
of the largest eddies (comparable to the size of the Pipe nebula $R_p$), 
gas drag would lead to a terminal relative velocity 
$v_r \sim (D_\rho \frac{R_c(m)}{R_p})^{\frac{1}{2}}v_g$ \citep{Lin2000} and
\eq
\tau_{\rm KH}\sim\frac{R_c(m)^{\frac{1}{2}}R_p^{\frac{1}{2}}}{v_g}
=1\ {\rm Myr}\left(\frac{R_c(m)}{0.1\ {\rm pc}}\right)^{\frac{1}{2}}\left(\frac{R_p}{10\ {\rm
pc}}\right)^{\frac{1}{2}}
\left(\frac{v_g}{1\ {\rm km\,s^{-1}}}\right)^{-1}
\label{eq:taokh2}
\ee

We include this ablation effect into the analysis of CMF's 
evolution as a source (due to the increase in the number of cores in the 
mass range $M-M_{BE}$) and sink terms (due to the decrease in the number of 
cores of mass M by ablation) such that
\eq
\left(\frac{df(m)}{dt}\right)_{\rm abla}=\int^{M_{\rm BE}}_{m}\frac{f(m')}{\tau_{{\rm KH},m'}}\frac{dm'}{m}
-\left(\frac{m-m_\mu}{m}\right)\frac{f(m)}{\tau_{{\rm KH},m}}.
\label{eq:ablation}
\ee
Kevin-Helmholtz instability would be 
suppressed by the cores' self gravity if their mass 
$M>M_{\rm BE}$ \citep{Murray1993}. They undergo gravitational 
contraction and collapse rather than ablation. The magnitude of 
$M_{\rm BE}$ is approximately 2 $M_\odot$ in the Pipe nebula case 
\citep{Lada2008}. This critical mass for ablation partly causes a 
peak in our simulated core mass distribution function.

\subsection{Other relevant processes}
In general, conduction between the warm inter-core gas and the smallest 
cores may lead to their evaporation into the inter-core gas. It is also 
possible for the inter-core gas to precipitate into new cores or 
condense onto existing cores.

We neglect the effect of precipitation which leads to the formation of 
lowest-mass cores. Condensation leads to a transfer of mass from the 
warm medium to cores at a rate ${\dot  M}_c = \pi R_c(m)^{2} c_s \rho_b$, 
where $c_s$ is the sound speed, $R_c(m)$ is the core radii for core with 
mass $m$ and $\rho_b$ is the background density. The associated 
conduction heats each core at a rate $H \sim {\dot  M}_c c_s^2$.
The cores are also cooling with rate $C= (4 \pi/3) R_c(m)^3 n^2 \Lambda$, 
where $n=\rho/(\mu M_H)$, $\mu$ and $M_H$ are the molecular weight and 
the mass of hydrogen atom. We compute the typical cooling rate 
$\Lambda$ following \citet{Bruce2011}, and test different evaporation 
rates ranging over 3 orders of magnitude. If heating exceeds cooling, 
the cores would evaporate on a time scale 
$M_c c_s^2 / (H-C)$. Otherwise, if $C>H$, the cooling of the gas inside 
the cores vanish as atoms in them reaches a ground state so that the 
condensation growth time scale would be $M_c c_s^2 / H$.

\section{Model}

\subsection{Observed properties of Pipe nebula}
We briefly summarize the observed properties of the Pipe nebula based 
on \citet{Lada2008}, \citet{GritschnederLin2012}:

\noindent
a) physical dimension of the Pipe nebula is 3 x 14 pc;

\noindent
b) overall mass of the Pipe nebula is $10^4$ \msun;

\noindent
c) total mass of the cores is $\sim 230$\msun (total mass is computed 
following the Table 2 of \citet{Rathborne2009});

\noindent
d) number density of inter-core gas is 774${\rm cm^{-3}}$, with mean 
molecular weight $\mu=1.37$ \citep{Lombardi2006}, gives average 
density of the inter-core gas is $1.77 \times 10^{-21}$g\,cm$^{-3}$;

\noindent
e) mean number density of core is $7.3\times10^3$ ${\rm cm^{-3}}$, 
with mean molecular weight $\mu=2.35$ \citep{Lombardi2006}, gives 
internal density of the cores is 
$\rho \simeq 2.87 \times 10^{-20}$ g\,cm$^{-3}$;

\noindent
f) size and mass range of cores are 0.05-0.15 pc and 0.2-20 \msun, 
respectively;

\noindent
g) Bonnor-Ebert mass of the cores: $M_{\rm BE} \simeq 2-3$ \msun;

\noindent
h) background gas velocity dispersion ($^{13}$CO line width) 
is $\sim1$ km\,s$^{-1}$ (see \citet{Onishi1999} Figure 6b, 
also \citet{Lada2008} \S 2.3.3;

\noindent
i) cores' non thermal velocity dispersion is $\sim 0.15$ km\,s$^{-1}$;

\noindent
j) typical sound crossing time in cores with $M=M_{\rm BE}$ is $\sim 1$ Myr; 
and

\noindent
k) dynamical time scale across the Pipe nebula is 3-10 Myr.

From these parameters, it is inferred that the pressure inside the core 
as well as in the inter-core gas are 
$P_{\rm int}/k \sim P_{\rm ext}/k \sim 10^5$ cm$^{-3}$ \citep{Lada2008}. 
We also find find the density contrast $D_\rho \simeq 15$, 
$\sigma \sim v_r \sim v_g \sim 1$ km s$^{-1}$, and $\tau_{\rm KH} \sim 1$ Myr.

\subsection{Numerical Setup}
Instead of investigating the evolution of individual cores, we evolve 
the mass distribution function with the processes described in \S\,2 
with the following set up.

In our standard model, we assume the warm medium is uniformly distributed 
with a total mass of $10^4 M_\odot$ within a sphere of radius $\sim5$ pc. 
The cores have an internal density of $7 \times 10^3$ cm$^{-3}$ so that 
their $M_{\rm BE}\sim 2 M_\odot$, which is consistent with that derived 
from the observation of starless clouds in the Pipe nebula \citep{Lada2008}.

The initial mass distribution of the cores is set up with a single 
power-law $dN/dlogM \propto M_c^{-\alpha}$. Analytical calculations predict 
the spectrum of non-self-gravitating structures to be 
$dN/dM \propto M^{2-n'/3}$, where $n'$ is the three dimensional power 
spectrum index of the log density field \citet{Hennebelle2008}. When 
taking $n'$ to be the Kolmogorov index $11/3$, we found $\alpha$ to be 0.67.
We only allow cores in the mass range between $M_{\rm min}$ 
(set to be 0.05 \msun) and 1 \msun at the onset of the calculation. 
Their total mass is set to be 250 \msun. For the standard model, we adopt 
$D_\rho\,=\,15$ and a 
constant (everywhere and throughout the calculation) transonic velocity 
dispersion ($\sim 1$ km\,s$^{-1}$) for all cores, regardless of their 
size (see Table \ref{t:numerical setup}). For a robustness test, we carried 
calculations with other values of model parameters such as 1) the 
cores' velocity dispersion, ranging from 
$0.5$\,km\,s$^{-1}$ to $2$\,km\,s$^{-1}$ 
2) and the slope of cores' initial mass distribution. 
The final result of the core distribution 
does not strongly depend on the above parameters in the range we tested.

The high-mass ($M_c > M_{\rm BE}$) cores are gravitationally unstable 
and they undergo gravitation contraction and eventually will evolve 
into stars. Based on an empirical model \citep{KrumholzTan}, we adopt a 
simplified prescription for the rate of star formation 
${\dot N} (M_c) = N (M_c)/\tau_\ast$. We set $\tau_\ast$ to be a 
$M_c$-independent characteristic time scale which is two orders of 
magnitude longer than their dynamical free-fall time scale. 
We do not account for mass loss and possible feedback to background mass 
during the star formation process. 
The star formation process contributes equally to all the cores exceed 
$M_{\rm BE}$. During the entire evolution the total mass of the stellar 
population is negligible compared with that of the cores.

We solve the combination of coagulation (Eq. \ref{eq:coag}) and 
ablation (Eq. \ref{eq:ablation}) equation as described in the appendix.
\eq
\frac{df(m)}{dt}=\left(\frac{df(m)}{dt}\right)_{\rm
coag}+\left(\frac{df(m)}{dt}\right)_{\rm abla}
\ee
The solution is then corrected with the effect of other processes 
(condensation, evaporation and star formation) in every time step. 
During their coagulation and ablation, the cores' mass is 
conservatively redistributed. The evaporation and condensation process
do not significantly modify the slope of core mass function in the most 
relevant mass ranges (0.3-10 \msun). It mainly changes the mass ratio 
between total core mass and the warm gas mass. While the mass fraction 
between warm medium and cold cores evolves with time, the total mass 
in the two components is essentially constant. The determining factors 
of core mass function are still dominated by coagulation and ablation over 
condensation, evaporation, and star formation.

\begin{deluxetable}{lccccccc}
\tablewidth{0pc}
\tablecaption{parameters in standard model\label{t:numerical setup}}
\tablehead{
\colhead{ $m_{\rm total}/m_{\rm totalc}$} & \colhead{$\rho_c $} &
\colhead{$d_{\rho}={\rho_c}/{\rho_b}$ } & \colhead{$dt$}
&\colhead{$m_{\mu}$} & \colhead{$m_{\rm max}$}  &  \colhead{$\sigma$}  &
\colhead{$M_{\rm BE}$} \\
}
\startdata
  $10^4 M_{\odot}$/250 $M_{\odot}$  & $3\times10^{-20}$ g cm$^{-3}$ & 15 &
10 yr &    0.05 $M_{\odot}$ &100 $M_{\odot}$  &  1 km s$^{-1}$    &  2
$M_\odot$
\enddata
\end{deluxetable}

\section{Result and Discussion}

The general shape of the observed CMF in the Pipe Nebula matches that of 
the IMF of young stellar clusters obtained by \citet{Kroupa2002b}. But 
the cores' mass at the peak of the CMF is around the Bonnor-Ebert mass, 
a factor of three larger than that of the stellar IMF (at least in the 
Pipe Nebula). We simulate models to reproduce the observed CMF.

In our model, we adopt a set of idealized initial conditions with a 
population of low-mass cores as described in \S 3.2. 
We found that the asymptotic mass function of the cores does not 
strongly depend on this choice of initial condition, albeit in the 
standard model, it takes $\sim 1$ Myr for the cores with 
$M_c>M_{\rm BE}$ to emerge and another $\sim 2-3$ Myr for the core 
mass function to establish a smooth distribution. Due to the low 
star formation efficiency in our model, stellar mass in the system 
is almost zero until around 5-6 Myrs, at which we terminated the 
evolution for not many propostellar cores have been found in Pipe yet.      

\begin{figure}
\centering
\includegraphics[width=0.6\linewidth]{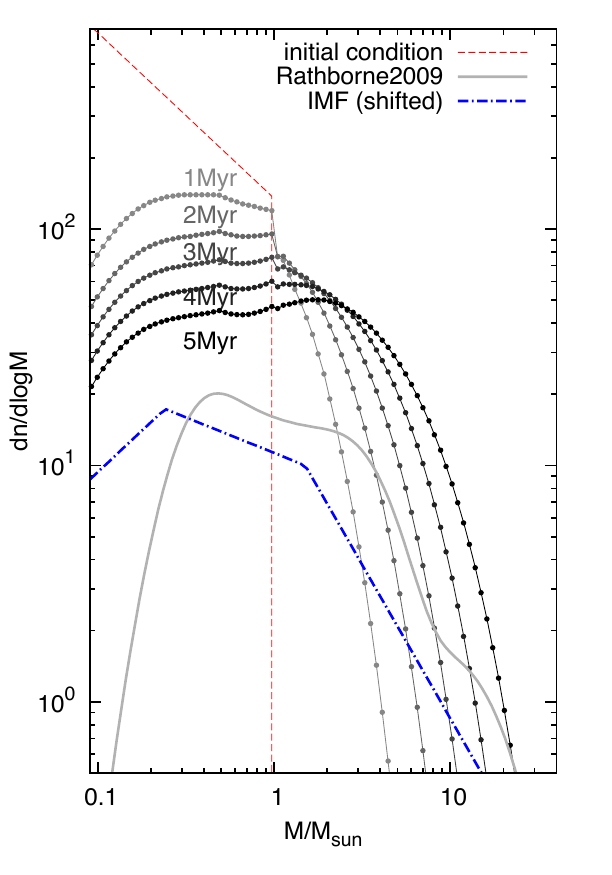}
\caption{
Red dash line show the initial distribution,
Black lines with dots from light to dark (end phase)
show the evolution of CMF within 5 Myr in equal time interval.
Blue doted-dash lines denote the Kroupa IMF, it is rescaled and shifted 
towards the high mass end by a factor of 3 for comparison. Grey line 
shows the CMF of Pipe Nebular as a probability density function 
from Figure 6 of \citet{Rathborne2009}. The lines are labeled with 
the evolution time from zero age.
\label{fig:evolution}
}
\end{figure}

We first compare the evolution of the simulated cores' mass function 
from the standard run with the observed CMF in the Pipe Nebula. The 
darkest black line in Figure \ref{fig:evolution} represents the stage 
after the emergence of gravitationally unstable cores but before the 
onset of star formation, as is observed in the Pipe Nebula today. 
For the mass range 0.3-10 $M_\odot$, this black line matches closely 
with the observed core distribution (grey line). We reproduced the 
broad peak of distribution around the Bonnor-Ebert Mass, a flat 
distribution around 1 $M_\odot$, and power-law cut offs at both low 
mass end and high mass end.
The mismatch at the low mass end $M_c<0.3M_\odot$ is partly due to 
observation bias as well as the simplified evaporation model we use.  
There are big uncertainties at the high mass end $M_c>10 M_\odot$, 
for in this region, both observation and our model need to deal with 
the problem of small number statistics.

We note that the cores at the high mass end of the CMF are 
gravitationally bound. Their free-fall time scale is typically less 
than 1 Myr. In order to match the shape of the CMF and accounts for 
the absents of stars in them, we adopt the assumption that they 
evolve on a time scale which is $\sim$ 30 Myr. In the case of Pipe, 
magnetic field $\sim 17-65 \mu\,G$ that might play an important role 
in the pressure buget is observed inside the 
nebula \citep{Alves2008}.
The existence of magnetic field will provide additional pressure 
support against gravity so that the cores might collapse on a much longer 
time scale compare to free-fall time. 

\begin{figure}
\centering
\includegraphics[width=0.8\linewidth]{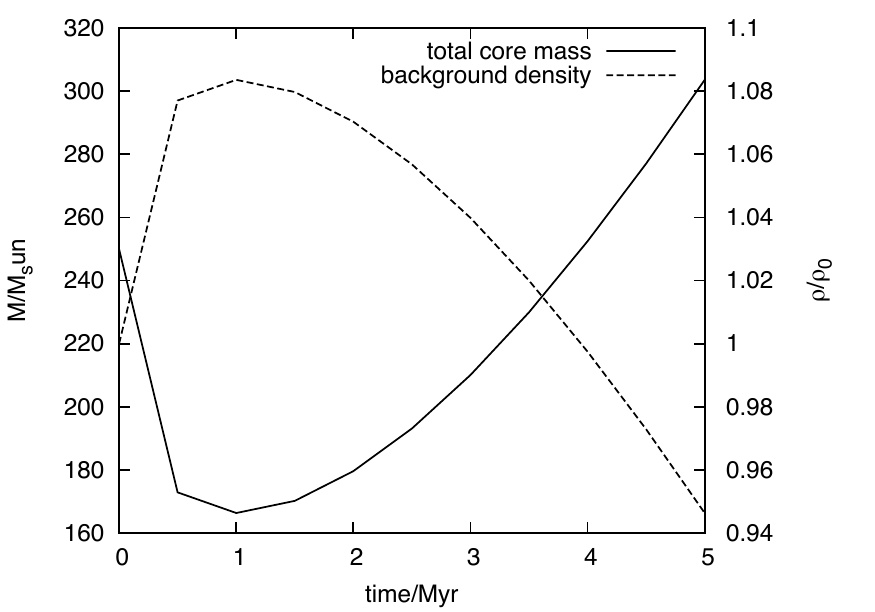}
\caption{
Evolution of total core mass (solid line) and 
warm gas density (dashed line) of the standard model.
\label{fig:evol}
}
\end{figure}

We show the total mass evolution of both components in Figure 
\ref{fig:evol}. During the early stage, when the total core mass is 
concentrated in small mass end, the effect of evaporation win over 
condensation, so that the total mass of core decrease in the first 
$1{\rm Myr}$. After the cores' mass function approaches from an initial 
arbitrary distribution to a broad self similar form, most of cores are 
not affected by evaporation anymore. The total mass of cores increase 
due to condensation. Since we terminated the evolution at the stage 
when stellar mass can be negligible, the background gas density 
show inverse evolution path compare to the total core mass. 

\begin{figure}
\centering
\includegraphics[width=0.8\linewidth]{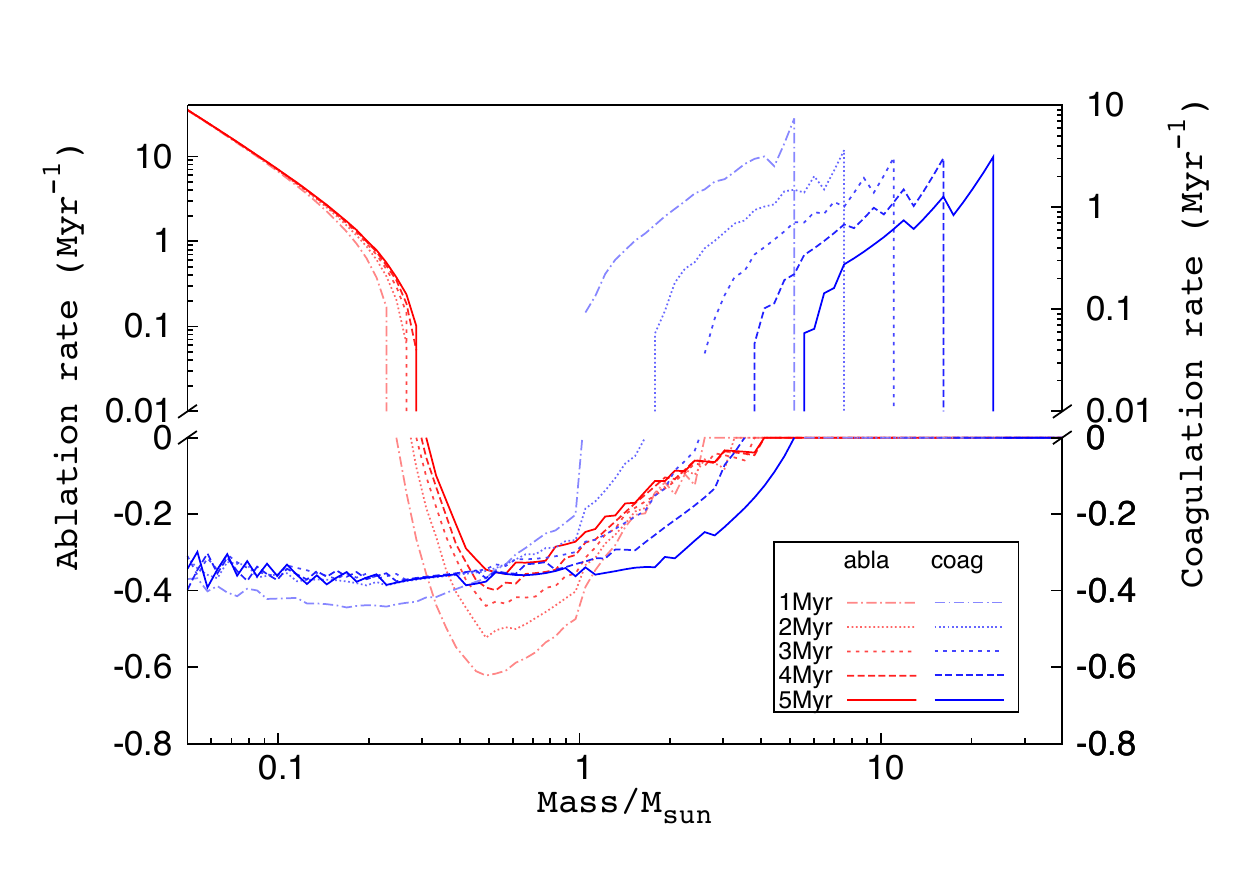}
\caption{
Comparison of ablation rate (red lines) and coagulation rate (blue lines) 
at different time of the evolution. The rate is computed with 
$(\frac{\delta\,n}{n\delta\,t})$ for every mass bin and for each 
process in units of ${\rm Myr^{-1}}$. $\delta\,n$ is the rate change due 
to an individual process in mass bin which has number of cores $n$, 
with time step $\delta\,t$. Note the upper half of the 
figure is in log scales and the scale for ablation rate and coagulation 
rate is different.
\label{fig:timescale}
}
\end{figure}

By comparing the evolution of rate of coagulation and ablation at 
different mass bins in Figure \ref{fig:timescale}, we can see the 
contribution of each processes to the final distribution.
In the low mass end, coagulation contributes as sink term and 
ablation contributes as source term. Although ablation always dominates 
coagulation in mass range smaller than 0.2 \msun, the cores of mass 
are small enough so that they are subjected to the influence of 
evaporation. Thus the overall population in this low mass range reduce 
during the evolution. From 0.2 \msun - 1 \msun, the number density 
decrease at almost constant rate since both coagulation and ablation 
act as sink terms. From 1-2 \msun, coagulation starts to contribute 
positively to the mass distribution. For massive cores, they 
are determined by a self similar solution from a coagulation equation 
only. As the system evolve, the rate of ablation increase due to 
decrease of background density, the peak of gain from coagulation 
moves towards higher mass along with the total distribution.  

There are some freedom in the detailed prescription of the
ablation process. In reality, 
the outer shell of a gravitationally bound Bonnor-Ebert sphere 
can still be stripped by KH instability. However, as shown in 
\citet{Murray1993}, the mass loss rate of core is really sensitive 
to the strength of gravitational field. Only $2\%$ of total mass 
got stripped by KH instability over 3.2 $\tau_{KH}$ in their 
simulation with a marginally gravitational bound cloud. 
In our model, for the gravitational bound clouds, the mass gain 
from coagulation will dominate the mass loss due to the stripping 
of surface layer, so that a strict terminate of KH process 
at $M_{BE}$ could be a good enough assumption.

Note that the time scale for transition from the initial to self similar 
mass distribution depends mostly on the initial filling factor and the 
velocity dispersion. The poorly known initial condition introduces 
additional uncertainties in the entire evolution timescale for the 
system. 

\begin{figure}
\subfigure[]
{
\centering
\includegraphics[width=0.5\linewidth]{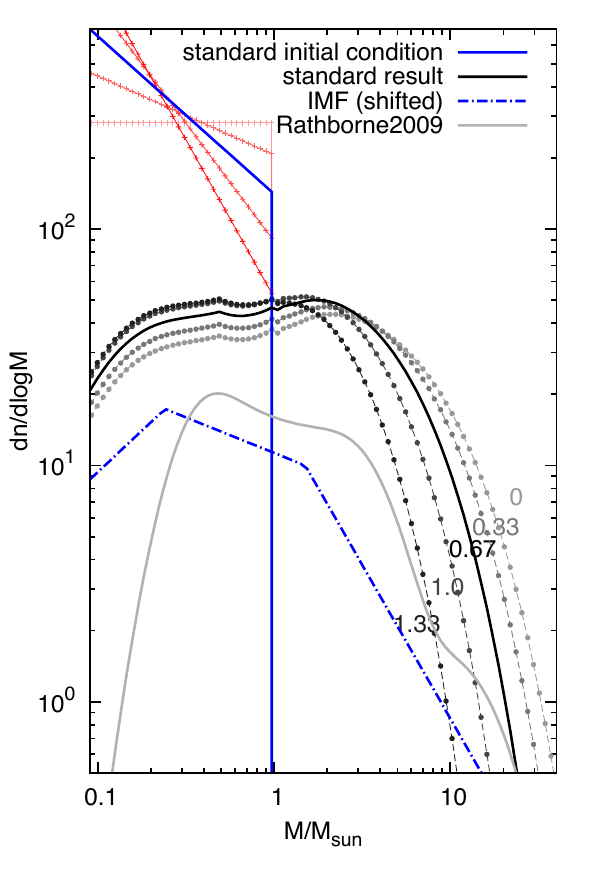}
\label{fig:Init}
}
\subfigure[]
{
\centering
\includegraphics[width=0.5\linewidth]{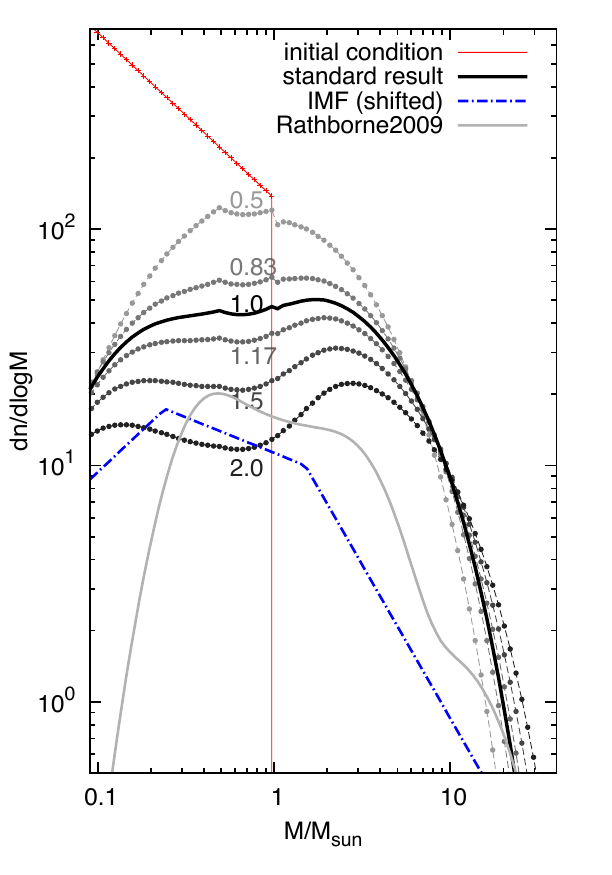}
\label{fig:Vc}
}
\caption{
The 5Myr CMFs result from different initial conditions. 
Blue dotted-dashed line in both figures is the rescaled and shifted 
Kroupa IMF; grey curve is the observed probability distribution 
of cores in Pipe nebular \citep{Rathborne2009}.
(a) We show the result of using different initial distribution 
of CMF at zero age. Red crosses (from light to dark) show the initial 
distributions for different runs. They were set up with formula 
$dN/dlogM\propto\,M^{-\alpha}$, while the power-law index $\alpha$ 
varying from 0-1.33; black dots with dash lines from light to dark 
show the corresponding final distribution of cores. 
Blue solid line show the initial distribution we use for the 
standard run, black solid line show the final distribution from 
the standard run. We label the result from different runs with 
the value of $\alpha$.
(b) We show the result of using different velocity dispersion 
(varying from 0.5 ${\rm km\,s^{-1}}$ - 2 ${\rm km\,s^{-1}}$) of the cores. Red solid line 
show the initial conditions shared by all the runs. Black dots 
with dash lines show the final distribution from different runs, 
from light to dark, the velocity dispersion increase. Solid black 
line is the result from the standard run, with a velocity dispersion 
of 1 ${\rm km\,s^{-1}}$. We label the result from different runs with the 
value of $\sigma$, in units of ${\rm km\,s^{-1}}$. 
\label{fig:permute}
}
\end{figure}

We explore a range of initial conditions in Figure \ref{fig:permute}. 
We vary one parameter at a time, with all the other set ups the same 
as the standard model. We terminate all the runs at 5Myr to make the 
comparison easier.
The left panel, Figure \ref{fig:Init}, presents the result (at 5Myr) 
from runs initialed with different core distribution at zero age. 
The power-law index of distribution $dN/dlogM\propto=M^{-\alpha}$ is 
varied from 0-1.33. As expected, a flatter initial distribution 
(lighter lines) assists coagulation so that the result distribution 
at 5Myr is shifted towards higher mass end. The initial slope does 
not strongly influence the shape of final distribution. In contrast, 
as shown in the right panel, Figure \ref{fig:Vc}, a different velocity 
dispersion ($\sigma$ varied from 0.5-2 ${\rm km\,s^{-1}}$) changes both 
the overall shape and the amount of mass in the massive cores. The slope 
of final distribution between $0.1-2 M_{\odot}$ is mainly due to the 
balance of coagulation and ablation. As we increase the velocity 
dispersion, the slope varies in this mass range. A higher velocity 
dispersion will also help with the emerge of the massive cores, lead 
to broader distribution with more mass in the high mass end. 
Nevertheless, in the allowed parameter space, the final core mass function is 
weakly dependent on initial conditions.  

In other star formation regions, the environment is somewhat different 
from the case of Pipe. Ophiushus has a much smaller global 
velocity dispersion $\sigma<0.4 {\rm km s^{-1}}$. If only consider 
coagulation and ablation, the evolution time scale will be longer, 
making them unlikely to be the only cause of the distribution. 
Other processes such as global gravitational contraction and 
accretion onto the core complex may also play important roles in 
the cores' velocity dispersion.
But as we shown in Figure \ref{fig:Vc}, within 5\,Myr, a system of 
velocity dispersion of $\sigma=0.5 {\rm km s^{-1}}$ can still 
evolve into a flat and broad peak around 1-2 \msun, demonstrating 
the two process could still contribute to the final distribution 
in such a system if the stellar less cores have longer lifetime due 
to the pressure support of either turbulence or magnetic field. 
Further more, the cores' size-dependent velocity 
dispersion is expected to have some influence on the power law 
index of CMF. This size dependence may originate from the cascade 
of turbulence in the warm gas and its ram pressure acceleration 
of the cores. Besides velocity dispersion, the diversity in 
Bonnor-Ebert mass in different regions is also likely to cause the 
diversity in the CMF.


\section{Conclusion}

In several nebula, including the Pipe, Aquila, Polaris and Ophiuchus,
a population of starless core have been observed, at the low mass end, 
they are confined by external pressure where as the most massive cores 
are bound by their own gravity. The origin of these cores mass function 
(CMF) is a crucial issue in understanding of star formation in these regions.

In this paper, we present our calculation based on a two-phase-media 
model. We show that the dominant physical processes in this environment 
are coagulation, ablation and thermal conduction. 
We demonstrate an IMF-like core mass function can be generated with a 
set of appropriate values. Through the exploration of a range of 
parameters reasonable for the physical condition of Pipe, we find this 
result is quite general.
This robust form of the CMF is an indication that it is determined by a 
dynamical equilibrium, {\it ie} a balance between two competing processes: 
ablation and coagulation.

The simulated CMF matches closely with the observed CMF in Pipe, and 
coincidentally matches the Kroupa IMF of the young stars.
This result suggests that the ablation clouds occur prior to their
collapse and the onset of star formation. The peak of the CMF is
established near the cores' Bonnor-Ebert mass mainly because ablation
is suppressed and collisional cross section is enlarged from their
physical size for cores with $M_c > M_{\rm BE}$. A similar peak is
observed in the stellar IMF, albeit with a corresponding stellar
mass which is one third that of the most populous cores. If the a few 
of these cores capture a large fraction of their masses, they would 
form massive stars with copious sources of ionizing photons. 
Their feedback on the background gas and the nearby cores may modify 
the Bonner-Ebert mass and lead to a reconciliation between the stellar 
IMF and the cores' CMF. We will explore these issues in our next 
contribution.

\section{Acknowledgements}
We thank Drs Matthias Gritschneder and Herbert Lau for useful conversation.
We also thank the anonymous referee for his/her valuable comments.
DNCL acknowledges support from NASA grant NNX08AL41G.

\appendix

\section{Implicit Method for solving Coagulation Equation and Ablation}

In general, the evolution of $f(m)$ can be determined by the
equation:

\begin{displaymath}
\frac{\frac{dN}{dm}}{dt}=\frac{df_j^i}{dt}\\
=\frac{1}{2}\int_{m_\mu}^{m_j-m_\mu}f_k^{i+1}f_{j-k}^{i+1}G_{k,j-k}\,dm_k\\
-\int_{m_\mu}^{m_{max}}f_j^{i+1}f_k^{i+1}G_{j,k}\,dm_k
\end{displaymath}
\eq
+\int^{m_{BE}}_{m_j}\frac{f_k^{i+1}}{\tau_{KH,k}}\frac{dm_k}{m_j}\\
-\left(\frac{m_j-m_\mu}{m_j}\right)\frac{f_j^{i+1}}{\tau_{KH,j}}
\ee
In the function, the upper superscript denotes for the time levels and the
lower superscript stands for different mass grids.
The first two terms are due to coagulation, and the last two terms
describe ablation. Change $m$ into non-dimensional $x$ ($x_j={m_j}/{M_\odot}$), and collect terms,
\begin{displaymath}
\frac{df_j^i}{dt}\\
=A\left[\int_{x_\mu}^{x_j-x_\mu}f_k^{i+1}f_{j-k}^{i+1}(x_k^\frac{1}{3}+x_{j-k}^\frac{1}{3})^2\\
\,dx_k\right.
\left.-2\int_{x_\mu}^{x_{\rm max}}f_j^{i+1}f_k^{i+1}(x_k^\frac{1}{3}+x_j^\frac{1}{3})^2\,dx_k\right]
\end{displaymath}
\eq
+C\left[\int^{x_{\rm BE}}_{x_j}\frac{f_k^{i+1}}{x_k^{\frac{1}{6}}}\frac{dx_k}{x_j}-\left(\frac{x_j-x_\mu}{x_j}\right)\frac{f_j^{i+1}}{x_j^{\frac{1}{6}}}\right]
\label{eq:ndim}
\ee
In Eq. \ref{eq:ndim}, we have
\eq
A=\frac{\pi}{2}f_f\frac{\sigma}{R_{c0}},\,\quad\quad
C=\tau_{\rm KH}^{-1}=\frac{\nu_f}{R_{c0}D_{\rho}^\frac{1}{2}}=\left(\frac{\nu_c}{R_{c0}R_b}\right)^{\frac{1}{2}}
\ee
while volume filling factor ({\bf $f_f$}) equals
\eq
f_f=\left(\frac{M_b\rho_c}{NM_c\rho_b}\right)^{-1}
\ee
$R_{c 0}$ is the radius of a 1~$M_\odot$ core. If assume $R_{c 0}$ invariant, and adopt $\nu_c$ and $D_\rho$ to be constants, C is
irrelevant to any other parameter. And A is only related to the volume
filling factor $f_f$
under this assumption, which is related to the thermal interaction between 
two phase media, requires further investigation in future work.

We solve Eq. \ref{eq:ndim} numerically implicitly with first order accuracy 
in time by discretize solution in both space and time. We use $N=100$ 
discrete mass bin ranging from minimum mass $m_{\mu}$ to maximum mass $m_{max}$ in logarithm scale.
(see Table (1), showing all chosen parameters.)
\eq
x_j=0.05*(1+0.09)^{j-1} \quad\quad(1<j<100)
\ee

Introducing in $\delta_j^i=f_j^{i+1}-f_j^i$, and ignoring the second order term of $\delta$,
we obtain a linear function of $\delta$.
\begin{displaymath}
\frac{\delta_j^i}{\Delta t}=A\left[\sum_{x_k=x_\mu}^{x_k=x_j-x_\mu}\\
(f_k^{i}f_{j-k}^{i}+\delta_k^{i}f_{j-k}^{i}+\delta_{j-k}^{i}f_k^{i})\\
(x_k^\frac{1}{3}+x_{j-k}^\frac{1}{3})^2\,\Delta x_k\right.
\end{displaymath}
\eq
\left.-2\sum_{x_k=x_\mu}^{x_k=x_{max}}[f_k^{i}f_{j}^{i}+\delta_k^if_{j}^{i}+\delta_{j}^if_k^{i}]\\
(x_k^\frac{1}{3}+x_j^\frac{1}{3})^2\,\Delta x_k\right]
+C\left[\sum^{x_k=x_{\rm BE}}_{x_k=x_j}
\left(\frac{\delta_k^i+f_k^{i}}{x_k^{\frac{1}{6}}}\right)\frac{\Delta \\
x_k}{x_j}-\left(\frac{x_j-x_\mu}{m_j}\right)
\left(\frac{\delta_j^i+f_j^{i}}{x_j^{\frac{1}{6}}}\right)\right]
\ee

which can be reduced to the form of
\eq
\sum[\Psi I_{j,k}+\Phi_{j,k}]\,\delta_{k}=B_j
\ee
Here, we take $I_{j,k}$ as an unit matrix, where $\Psi$ {\bf and $\Phi$} are 
some constant only related to mass distribution in the old time level.

Instead of implementing a thoroughly numerical analysis for the stability properties of 
explicitly method on Eq \ref{eq:ndim}, we use the result from this implicit 
method  to justify the robustness of the solution. For reasonable time steps 
and initial conditions, we are also able to get  consistent result from 
explicit numerical method. To reduce numerical time cost, we'll switch to 
explicit method for further parameter space investigation in verified region.


\end{document}